# Evaluation of Competing J domain:Hsp70 Complex Models in Light of Existing Mutational and NMR Data

Rui Sousa*, Jianwen Jiang, Eileen M. Lafer, Andrew P. Hinck,
Liping Wang, Alexander B. Taylor, and E. Guy Maes

Dept. of Biochemistry, U. of TX Health Sci. Ctr.,  7703 Floyd Curl Drive, San Antonio TX 78229-3900
*Correspondence: sousa@biochem.uthscsa.edu, 210-567-2506, 210-567-6595 (fax)

Ahmad et al. recently presented an NMR-based model for a bacterial DnaJ J domain:DnaK(Hsp70):ADP complex(1) that differs significantly from the crystal structure of a disulfide linked mammalian auxilin J domain:Hsc70 complex that we previously published(2).  They claimed that their model could better account for existing mutational data, was in better agreement with previous NMR studies, and that the presence of a cross-link in our structure made it irrelevant to understanding J:Hsp70 interactions.  Here we detail extensive NMR and mutational data relevant to understanding J:Hsp70 function and show that, in fact, our structure is much  better able to account for the mutational data and is in much better agreement with a previous NMR study of a mammalian polyoma virus T-ag J domain:Hsc70 complex than is the Ahmad et al. complex, and that our structure is predictive and provides insight into J:Hsp70 interactions and mechanism of ATPase activation.

**I. Comparison with previous NMR studies of J-Hsp70 interactions**:  The regions of the J domains of 2 J proteins that interact with Hsp70 have been previously mapped by NMR chemical shift analysis.  Greene et al.(3) mapped them in a complex of bacterial DnaJ and DnaK and Garimella et al.(4) mapped them in a complex of polyoma virus T-ag J domain and bovine Hsc70.  Ahmad et al. state that the data from their complex agrees with the data of Greene et al. and that our structure does not.  The Ahmad et al. complex does show better agreement with the Greene et al. study, though there are discrepancies: Greene et al. observe shifts in the functionally critical H and D residues of the invariant HPD motif, while Ahmad et al. conclude that their own data do not support a role for the HPD loop in binding DnaK:ADP, and say that they observed "no chemical shift changes for residues Asp35 (and His33) of the HPD loop…It appears that the HPD motif is exclusively involved in the interaction with the ATP state, although it could also have a purely structural role." Ahmad et al. suggest that the Greene et al. results for D35 could have been due to asp-specific pH effects.  However Greene et al. observed significant peak broadening for H33 as well as D35 upon titration with DnaK:ADP so asp-specific pH effects or differences in nucleotide state cannot explain this discrepancy.

More importantly, our structure is of a *mammalian* Hsc70:J complex, and specifically we use bovine Hsc70, as was used by Garimella et al. When our structure is compared to the chemical shift/peak broadening/protection data from Garimella et al. we find excellent agreement: 8 of 13 residues in PyJ identified as shifted/broadended by Garimella et al. correspond to auxilin residues that are close to Hsc70 in our structure.  In contrast, the Ahmad et al. complex is entirely inconsistent with the Garimella data: none of the residues identified as broadened/shifted in Garimella correspond to J residues identified in the Ahmad et al. complex as close to labeled DnaK residues.  This is expected.  The Garimella and Greene studies indicated that prokaryotic Hsp70 binds DnaJ differently than how mammalian Hsc70 binds J protein.  Bacterial DnaJ uses primarily J domain helix II to bind bacterial DnaK, while the mammalian (viral) protein uses primarily helix III to bind mammalian Hsc70.  The different binding mode of Hsc70:PyJ vs. DnaK:DnaJ was *the* central conclusion of the Garimella study, being incorporated in the title, addressed in the abstract ("our novel evidence implicating helix III differs from evidence that Escherichia coli DnaK primarily affects helix II and the HPD loop of DnaJ."), and in the introduction, results, and discussion of that study.  This conclusion is also supported by extensive mutational data (see below).  It is primarily auxilin J domain helix III which contacts Hsc70 in our structure.  Since our complex is of a mammalian Hsc70 and a mammalian J, it should be compared, and is expected to be more similar to, the mammalian complex studied by Garimella, rather than the bacterial complex studied by Greene.  The excellent agreement between the Garimella et al. results and our structure is strong evidence for the physiological relevance of the latter.   These NMR data and relevant references are summarized in table 1.

**II. Mutational data:**

1. At least 38 single, multiple point and deletion mutants in the bacterial DnaJ J domain have been characterized for complementation in vivo, with a subset also characterized for binding to and ability to stimulate DnaK in vitro(*5-7*). These mutations identify 9 residues (Y25, R26, H33, P34, D35, R36, N37, F47) that result in complementation defects when mutated. Mutants of H33 and D35 have also been shown to disrupt binding to DnaK in vitro(*6, 8*). Of these 9 residues only one (K26) is identified as within 5-15 Å of one of the 6 DnaK residues in the Ahmad et al. complex that were spin-labeled to provide distance constraints. There are another 5 residues identified in this complex as being within 5-15 Å of a labeled DnaK residue. Mutations at these 5 residues do not give rise to complementation defects. In contrast, in our structure, there are 27 auxilin J domain residues close (<8 Å) to the Hsc70 NBD. These close residues correspond to 8 of the 9 functionally critical DnaJ residues.

2. The more relevant mutational data sets for evaluating our structure, however, are not those of bacterial DnaJ, but of viral T-ag J domains which bind the mammalian Hsc70 we used to prepare our complex. As noted above, NMR studies indicate that mammalian Hsc70 binds J very differently than how bacterial DnaK binds DnaJ. The mutational data sets(*9-11*) on the viral T-ag J domains are too extensive to fully summarize (one study(*9*) examined 63 mutants in polyoma virus large T-ag J domain and 51 in middle T J domains). However, there are at least 14 well-expressed point mutations in polyoma or SV40 large T-ag J domains shown to disrupt complementation *in vivo* or binding/ATPase stimulation *in vitro*. None of these mutations are at residues identified in the Ahmad et al., complex as within 15 Å of labeled residues in DnaK, and 5 are more than 20 Å away. In contrast, 12 of the 14 corresponding auxilin residues are within 8 Å of Hsc70 in our complex.

3. On the DnaK NBD there are at least 10 single or 2-4 residue segments that affect J binding or ATPase stimulation (Y146-D148, R151, D388, D393, R167, N170, T173, E217V218, V388-L392, L390L391)(*6, 12-14*). Where it has been examined, mutations of corresponding residues in bovine Hsc70 are also found to affect J binding/stimulation(*15*) and experiments with bovine Hsc70 identify 2 additional residues (I216 and L380 corresponding, respectively, to DnaK T215 and T383)(*2*) that abolish J stimulation and binding. The Ahmad et al. complex identifies 16 DnaK residues (206-211) that are within 5-15 Å of one of the 3 labeled DnaJ residues. Of these residues only 2 (217,218) correspond to a region on DnaK that affects interaction with J when mutated. The other 11 functionally critical residues/regions in DnaK are all more than 20 Å from the labeled DnaJ residues in the Ahmad et al. complex. The Ahmad et al. complex also identifies a close approach of J to DnaK residues 208/209, but mutation of these residues does not abrogate J stimulation/binding (see point 5 below). In contrast, in our structure, there are 33 Hsc70 NBD residues close (<8 Å) to the auxilin J domain. These close residues encompass 10 of the 12 residues/regions that affect J binding/stimulation when mutated.

All of this published mutational data is summarized in Tables 2-4, along with relevant references.

4. Ahmad et al. state that the cysteine mutants used to introduce spin labels did not affect ATP hydrolysis, "these (and all other spin-labeled mutants reported upon here) were unperturbed in the ATP-hydrolysis activity assays (Fig. S5)". This is a misrepresentation. Inspection of figure S5 shows that all of these mutations have effects on ATP hydrolysis, with reductions in initial rates of 2-10 fold.

5. Ahmad et al. state that the effects of mutating DnaK DE208,209 support their proposed role for these residues in binding J "DnaK DE(208,209) AA is only partially stimulated by DnaJ, whereas the ATP hydrolysis of DnaK DE(208,209)AA by itself is not affected (Fig. S6)." This is a misrepresentation. Inspection of figure S6 reveals that the J stimulated ATPase rate of DE(208,209)AA is fully ~80% of WT while the unstimulated rate of this mutant is ~30% greater than the WT. In addition, we have characterized Hsc70 mutations that correspond to DnaK ED208,209 (ED213,214AA). This mutant is WT in its ability to bind and be stimulated by auxilin and to mediate auxilin dependent clathrin coat disassembly(*15*).

6. Ahmad et al. state that the identified J interacting region (aa 206-221) on DnaK is well conserved and that this may explain the cross-species functionality of the J domain. The nucleotide binding domains of Hsp70s are one of the best conserved protein families known so almost any region is likely to show some cross-species conservation but, in fact, the DnaK 206-221 loop, overall, is one of the more poorly conserved regions in the Hsp70 NBD, with extensive variation at residues 208/209, a deletion of residue 210 (in K relative to other 70s) and a 4 residue insertion of residues 211-215 in E. coli DnaK that is not present in the vast majority of Hsp70s(*16*).

### III. Functional Relevance of the Ahmad et al. Complex and our Structure.

1. The relevance of the J domain:Hsp70 complex studied in the ADP state by Ahmad et al. is unclear. Ahmad et al. estimates the $K_d$ for DnaJ binding to their DnaK:ADP:substrate complex to be ~16 µM, which would make it too weak to be physiogically relevant. However, its possible that it plays a transient role during protein substrate handoff when the J protein and Hsp70 protein substrate binding domains both engage different regions of the same protein substrate. Though biochemical studies have indicated that the J domain:Hsp70 interaction is stable only in the ATPase state(*6, 17-19*), is possible that the high concentrations of this protein achieved in an NMR tube allow a weak J domain:DnaK ADP state interaction to be detected. We would note, however, that we have been unable to detect chemical shifts with Hsc70 NBD and auxilin J domain even at >[100 µM] of the latter (unpublished), so it may be that the auxilin J:Hsc70:ADP interaction is even weaker than the Dna J:DnaK:ADP interaction. We therefore prepared the cross-linked auxilin J domain:Hsc70 complex with the aim of gaining insight into the structure of the *ATP* state complex. Indeed, we found that the presence of ATP in the crosslinking reaction enhanced the formation of the cross-linked species. Ahmad et al. acknowledge that the ATP and ADP state complexes may differ, but does not acknowledge the possibility that differences between his ADP state complex and ours may be due to the fact that our complex resembles the ATP state.

2. Ahmad et al. state that our complex cannot be physiologically relevant because it contains a disulfide cross-link. However, the engineering of cross-links to allow crystallization of transient or excessively dynamic complexes was not unprecedented when we did it, and has become a fairly common approach. We searched the PDB with the terms "disulfide linked", "crosslinked", "crosslink","disulfide-linked", "cross-linked", and "cross-link" and identified 558 entries. Not all of these hits represent crosslinks engineered for structure determination, but sampling of these entries indicates that a large fraction are (time precluded an inspection of all 558 entries to obtain precise statistics). Examples would be the Verdine and Harrison groups' preparation of HIVI RT:DNA co-crystals (1RTD(*20*)), of the Steitz groups' preparation of DNAP:DNA co-crystals (1KLN(*21*)), or of the Poulos' group determination of a cross-linked cytochrome c peroxidase-cytochrome c structure (1S6V(*22*)). These approaches allow determination of the structure of a transient or dynamic complex that cannot otherwise be captured, and while the crosslink restricts conformational freedom at the crosslink site, it does not dictate the conformation of the complex as a whole. In the complex we prepared there is freedom of rotation around the disulfide bond so the auxilin J domain is free to settle into the most energetically favorable configuration on the surface of the Hsc70 and its position in the crystal structure is informative. It is, however, critical that other evidence support the physiological relevance of such a complex. In our case we showed that(*2*): i. In solution, the auxilin J domain stimulated the ATPase activity of a construct encompassing the Hsc70 nucleotide binding domain and interdomain linker (NBD_Linker) but not of the NBD alone. The complex precisely recapitulated this solution behavior: the crosslinked auxilin stimulated ATP hydrolysis by the NBD_Linker but not by the NBD alone. In addition, ATP enhanced crosslinking to the NBD_Linker but not the NBD, ii. The complex was in excellent agreement with previous mutational studies, particularly for the viral T-ag J domains, and with the Garimella NMR study of the polyoma virus T-ag J domain:Hsc70 complex, as detailed above, iii. The complex successfully predicted that mutation of L380 would abrogate auxilin's ability to stimulate Hsc70 ATPase activity, iv. The complex revealed that the cross-linked auxilin induced partial ordering of the interdomain linker and suggested how the J domain could stimulate ATPase activity by conformational effects on the linker, a mechanism that is gaining increasing support(*23*), v. The complex is consistent with the cryo-EM structure of a clathrin coat:Hsc70(ATP):auxilin complex(*24*).

**Summation:**

1. Our cross-linked bovine Hsc70:auxilin J domain complex is in excellent agreement with the NMR study of the bovine Hsc70:polyoma virus T-ag J complex, but not in agreement with the NMR study of the bacterial DnaJ:DnaK complex. This is expected since it is known that the T-ag J complex and the bacterial complex bind via different modes and we expect the two mammalian complexes to be more similar to each other than to the bacterial complex. Consistent with this, the Ahmad et al. complex shows no agreement with the T-ag J complex and agreement with the previous NMR study of DnaJ:DnaK, though there are surprising discrepancies with the latter, including the unexpected observation that the HPD loop is not involved in binding DnaK.

2. Our structure can better account for existing mutational data: the large majority of residues on the Hsp70 or J domain surface that have been shown to affect Hsp70:J domain interaction map to residues at the interprotein interface in our structure (8/9 for DnaJ; 12/14 for Py or SV40 J;10/12 for Hsp70). In the Ahmad et al. complex only a minority of these residues are indicated to be at the interprotein interface (1/9 for DnaJ; 0/14 for Py or SV40 J; 1/12 for Hsp70).

3. Our cross-linked structure recapitulates the solution behavior of the non cross-linked species, has been shown to predict residues whose mutation affects J:Hsc70 interactions, and suggests a mechanism for J stimulation of ATPase activity through effects on the interdomain linker consistent with evolving models.

We would have carried out a more thorough inspection of the Ahmad et al. complex if we could have but there was no statement of coordinate deposition in the PDB or accession number given. Ahmad et al. stated that such coordinates were in the supplemental data: "The coordinates of 50 MD snapshots for the DnaJ–DnaK complex are given in the SI Text.", but when we examined the supplemental data we could find no such coordinate set.

| PyJ residues shifted/broadened/ protected (Garimella(*4*)) (Ahmad et al., complex)* | DnaJ residues shifted or broadened (Greene(*3*)) | Corresponding residue and distance to Hsc70 in Jiang(*2*) complex | | Closest NMR derived distance constraint to labeled DnaK |
|---|---|---|---|---|
| | Y6[@] | N829 | >8 | 20-200 |
| | V12[#] | M853 | >8 | 15-20 |
| | S13 | A854 | >8 | 20-200 |
| | A16[#] | V857 | >8 | 20-200 |
| | R19[#] | E860 | >8 | 15-20 |
| | E20 | Q861 | >8 | 15-20 |
| | I21 | V862 | >8 | 20-200 |
| Q32 | | K864 | >8 | 15-20 (K23) |
| | A24 | V865 | >8 | 15-20 |
| | Y25[#] | Y866 | **<8** | 15-20 |
| | K26 | R867 | >8 | **5-15** |
| | R27 | K868 | >8 | **5-15** |
| | L28 | A869 | >8 | **5-15** |
| | M30 | L871 | >8 | **5-15** |
| | Y32 | V873 | **<8** | **5-15** |
| | H33 | H874 | **<8** | 15-20 |
| | D35 | D876 | **<8** | 15-20 |
| D44 | | K877 | **<8** | 15-20 (R36) |
| K45 | | K877 | **<8** | 15-20 (R36) |
| A50 | | M889 | **<8** | 20-200 (A45) |
| L51 | | I890 | **<8** | 20-200 (K46) |
| M52 | | F891 | **<8** | 20-200 (F47) |
| Q53 | | M892 | **<8** | 20-200 (K48) |
| N56 | | N895 | **<8** | 15-20 (K51) |
| | Y54[@] | W898 | **<8** | 15-20 |
| G60 | | S899 | **<8** | 15-20 (E55) |
| T61 | | E900 | >8 | 15-20 (V56) |
| K63 | | E902 | >8 | 20-200 (T60) |
| T64 | | N903 | >8 | 20-200 (D61) |
| E65 | | Q904 | >8 | 20-200 (S62) |

Table 1. Residues in the polyoma virus (Py) T-ag or DnaJ J domain that are shifted/broadened/protected upon binding to, respectively, bovine Hsc70 or DnaK and their (or corresponding residue) proximity to Hsc70 or DnaK in, respectively, the Jiang or Ahmad et al. complexes. [@]Shifted/broadened in ADP state only. [#]Shifted/broadened in ATP state only. *Distances are to the nearest spin-labeled K residue (V210 which is in the center of the loop proposed by Ahmad et al., to form the DnaJ binding site on DnaK) as given in table S3 in ref. 1. Close approaches are bolded; DnaJ residues corresponding to the PyJ or SV40 J residues are given in parenthesis.

| J Mutation | Effect on J function | Ref. | NMR derived distance constraints (Å; Ahmad et al.)* | Distance (Jiang et al. complex) |
|---|---|---|---|---|
| $E_8A$ | none | 5 | V210/D326/D148/K166/T417/K421: 20-200 | >8 |
| $S_{13}A$ | none | 5 | V210: 15-20; D326/D148/K166/T417/K421: 20-200A | >8 |
| $K_{14}T_{15}AA$ | none | 5 | V210/D326/D148/K166/T417/K421: 20-200A | >8 |
| $E_{17}A$ | none | 5 | V210: 15-20; D326/D148/K166/T417/K421: 20-200A | >8 |
| $E_{18}A$ | none | 5 | V210: 15-20; D326/D148/K166/T417/K421: 20-200A | >8 |
| $R_{19}E_{20}AA$ | none | 5 | V210: 15-20; D326/D148/K166/T417/K421: 20-200A | >8 |
| $R_{22}K_{23}AA$ | none | 5 | V210: 15-20; D326/D148/K166/T417/K421: 20-200A | >8 |
| **$Y_{25}A$** | **complementation defect** | 5 | V210: 15-20; D326/D148/K166/T417/K421: 20-200A | **≤8** |
| **$K_{26}A$** | **complementation defect** | 5 | **V210: 5-15;** D326/T417: 15-20; K166/K421: 20-200A | >8 |
| $R_{27}A$ | none | 5 | **V210: 5-15;** D326/T417: 15-20; K166/K421: 20-200A | >8 |
| $L_{28}A$ | none | 5 | **V210: 5-15;** D326/T417: 15-20; K166/K421: 20-200A | >8 |
| $A_{29}G$ | none | 5 | **V210: 5-15;** D326/T417: 15-20; K166/K421: 20-200A | >8 |
| $M_{30}K_{31}AA$ | none | 5 | **V210: 5-15;** D326/T417: 15-20; K166/K421: 20-200A | >8 |
| $Y_{32}A$ | none | 5 | V210/D326/D148/K166/T417/K421 | **≤8** |
| **$H_{33}Q$** | **complementation defect** | 5 | V210: 15-20; D326/D148/K166/T417/K421: 20-200A | **≤8** |
|  | **K binding defect** | 7 | V210: 15-20; D326/D148/K166/T417/K421: 20-200A | **≤8** |
| **ΔH$_{33}$** | **complementation defect** | 5 | V210: 15-20; D326/D148/K166/T417/K421: 20-200A | **≤8** |
| **ΔP$_{34}$** | **complementation defect** | 5 | V210: 15-20; D326/D148/K166/T417/K421: 20-200A | **≤8** |
| **P$_{34}$F** | **complementation defect** | 5 | V210: 15-20; D326/D148/K166/T417/K421: 20-200A | **≤8** |
| **ΔD35** | **complementation defect** | 5 | V210: 15-20; D326/D148/K166/T417/K421: 20-200A | **≤8** |
| **D$_{35}$N** | **Abolishes binding to WT Suppresses K R167H** | 6 | V210: 15-20; D326/D148/K166/T417/K421: 20-200A | **≤8** |
| **R$_{36}$G** | **complementation defect** | 5 | V210: 15-20; D326/D148/K166/T417/K421: 20-200A | **≤8** |
| **N$_{37}$G** | **complementation defect** | 5 | V210/D326/D148/K166/T417/K421: 20-200 | **≤8** |
| $Q_{38}G$ | none | 5 | V210/D326/D148/K166/T417/K421: 20-200 | **≤8** |
| $G_{39}D_{40}$ | none | 5 | V210/D326/D148/K166/T417/K421: 20-200 | **≤8** |
| $K_{41}E_{42}$ | none | 5 | V210/D326/D148/K166/T417/K421: 20-200 | **≤8** |
| $E_{44}A$ | none | 5 | V210/D326/D148/K166/T417/K421: 20-200 | **≤8** |
| $K_{46}A$ | none | 5 | V210/D326/D148/K166/T417/K421: 20-200 | **≤8** |
| **F$_{47}$A** | **complementation defect** | 5 | V210/D326/D148/K166/T417/K421: 20-200 | **≤8** |
| $K_{48}E_{49}AA$ | none | 5 | V210: 15-20; D326/D148/K166/T417/K421: 20-200A | **≤8** |
| $K_{51}E_{52}AA$ | none | 5 | V210: 15-20; D326/D148/K166/T417/K421: 20-200A | **≤8** |
| $Y_{54}A$ | none | 5 | V210: 15-20; D326/D148/K166/T417/K421: 20-200A | **≤8** |
| $E_{55}A$ | none | 5 | V210: 15-20; D326/D148/K166/T417/K421: 20-200A | **≤8** |
| $T_{58}A$ | none | 5 | V210: 15-20; D326/D148/K166/T417/K421: 20-200A | >8 |
| $T_{58}D_{59}AA$ | none | 5 | V210/D326/D148/K166/T417/K421: 20-200 | >8 |
| $S_{60}Q_{61}AA$ | none | 5 | V210/D326/D148/K166/T417/K421: 20-200 | >8 |
| $K_{62}R_{63}AA$ | none | 5 | V210/D326/D148/K166/T417/K421: 20-200 | >8 |
| $D_{66}Q_{67}AA$ | none | 5 | V210/D326/D148/K166/T417/K421: 20-200 | >8 |
| Δ58-69 | none | 5 | V210/D326/D148/K166/T417/K421: 20-200 | >8 |

Table 2: Effects of DnaJ mutations on function and the proximity of the mutated (or corresponding) residues to DnaK or Hsc70 in, respectively, the Ahmad et al. vs. Jiang et al. complex. *Distance constraints are from the J domain residues to the spin labeled DnaK residues specified in column 4. For the Jiang complex we chose 8 Å as the distance limit for residues designated as 'close', as such residues are either in contact with Hsc70 or could move into contact given minor adjustments in the complex upon reduction of the disulfide link or relaxation in crystal packing.

| Mutations affecting J function* | Ref. | Closest NMR distance constraint (Å; Ahmad et al.,) | Distance (Jiang et al, complex) |
|---|---|---|---|
| Py_Q32A(K23) | 9 | 15-20 | >8 |
| Py_A33G (A24) | 9 | 15-20 | >8 |
| Py_Y34F (Y25) | 9 | 15-20 | <8 |
| **Py_P43S (P34)** | 9 | 15-20 | **<8** |
| **Py_H49R (A45)** | 9 | 20-200 | **<8** |
| **Py_M52V (K48)** | 9 | 20-200 | **<8** |
| **Py_N56T (K51)** | 9 | 20-200 | **<8** |
| SV_Y34N (Y25) | 10 | 15-20 | <8 |
| **SV_H42R (H33)** | 10 | 15-20 | **<8** |
| **SV_P43F (P34)** | 11 | 15-20 | **<8** |
| **SV_D44N (D35)** | 11 | 15-20 | **<8** |
| **SV_K45N (R36)** | 11 | 15-20 | **<8** |
| **SV_G47E (A43)** | 11 | 20-200 | **<8** |
| **SV_K53R (K48)** | 10 | 20-200 | **<8** |

*Table 3. Mutations in Polyoma (Py) or SV40 (SV) virus T-ag J domain that affect J function and the proximity of the corresponding residues in DnaJ or auxilin to DnaK or Hsc70, respectively. Bolded and underlined mutations are in the HPD loop. Bolded only correspond to residues in helix III of the J domain and unbolded to helix II.

| K Mutation | Effect on K function | Ref. | NMR derived distance constraints (Å; Ahmad et al.) | Distance (Jiang et al. complex) |
|---|---|---|---|---|
| $N_{147}A$ | 2-fold weaker binding to J | 6 | R19/M30/K41: 20-200 | >8 |
| $D_{148}A$ | WT binding to J | 6 | R19/M30/K41: 20-200 | >8 |
| $R_{151}A$ | **Reduced J stimulation of ATPase** | 13 | R19/M30/K41: 20-200 | >8 |
| $Q_{152}A$ | WT binding to J | 6 | R19/M30/K41: 20-200 | >8 |
| $K_{155}D$ | Near wt J stimulation of ATPase | 13 | R19/M30/K41: 20-200 | >8 |
| $R_{167}H$ | **7-fold weaker binding to J suppresses J D35N** | 6 | R19/M30/K41: 20-200 | <u>≤8</u> |
| $R_{167}A/D$ | **Reduced J stimulation of ATPase** | 13 | R19/M30/K41: 20-200 | <u>≤8</u> |
| $I_{169}A$ | WT binding to J | 6 | R19/M30/K41: 20-200 | <u>≤8</u> |
| $N_{170}A$ | **9-fold weaker binding to J** | 6 | R19/M30/K41: 20-200 | <u>≤8</u> |
| $T_{173}A$ | **11-fold weaker binding to J** | 6 | R19/M30/K41: 20-200 | <u>≤8</u> |
| $Q_{378}A$ | WT binding to J | 6 | R19/M30/K41: 20-200 | |
| $Y_{146}ND_{148}AAA$ | **Reduced ATPase stim by J Comp. defect, reduced refolding** | 14 | R19/M30/K41: 20-200 | >8 |
| $E_{217}V_{218}AA$ | **Reduced ATPase stim by J Comp.defect, reduced refolding** | 14 | **M30: 5-15**, /R19/ K41: 20-200 | <u>≤8</u> |
| $V_{389}LLL_{392}AAA$ | **Eliminates ATPase stim by J** | 15 | R19/M30/K41: 20-200 | <u>≤8</u> |
| $L_{390}L_{391}DD$ | **Eliminates ATPase stim by J** | 15 | R19/M30/K41: 20-200 | <u>≤8</u> |
| $D_{388}R$ | **Enhanced J stimulation of ATPase** | 15 | R19/M30/K41: 20-200 | <u>≤8</u> |
| $D_{393}A/R$ | **Reduced J stimulation of ATPase** | 15 | R19/M30/K41: 20-200 | ≤8 |
| *$I_{216}T$ (K $T_{215}$) | **Abolishes J binding and stimulation of ATPase** | 2 | R19/M30/K41: 20-200 | ≤8 |
| *$E_{213}D_{214}AA$ (K $D_{208}E_{209}$) | WT binding/stimulation by J | 16 | **M30: 5-15**, /R19/ K41: 20-200 | <u>≤8</u> |
| *$L_{380}G$ (K $T_{383}$) | **Abolishes J stimulation of ATPase** | 2 | R19/M30/K41: 20-200 | <u>≤8</u> |
| *$V_{388}C$ (-) | **Reduces J binding and inverts J stimulation of ATPase** | 2 | R19/M30/K41: 20-200 | <u>≤8</u> |
| *$L_{393}C$ (K $L_{390}$) | **Reduces J binding and inverts J stimulation of ATPase** | 2 | R19/M30/K41: 20-200 | - |

Table 4. Mutations in DnaK or bovine Hsc70 that affect interaction with the J protein and their proximity to J in either the Ahmad et al. or Jiang et al. complex. *Mutations in bovine Hsc70. The corresponding DnaK residue is given in parenthesis.